\documentclass[]{spie}  

\newcommand{\es}[2]{#1\times10^{#2}}
 
\usepackage{amsmath,amsfonts,amssymb}
\usepackage{graphicx}
\usepackage[colorlinks=true, allcolors=blue]{hyperref}

\title{Calibrations for high precision differential astrometry onboard Theia and HWO}

\author[a,b,c]{Manon Lizzana}
\author[a]{Fabien Malbet}
\author[a]{Hugo Rousset}
\author[a]{Sébastien Soler}
\author[a]{Fabrice Pancher}
\author[d]{\'Eric Thi\'ebaut}

\affil[a]{Univ.\ Grenoble Alpes, CNRS, IPAG, F-38000 Grenoble, France}
\affil[b]{CNES, F-75001 Paris, France}
\affil[c]{Pyxalis, F-38430 Moirans, France}
\affil[d]{Université Lyon 1, ENS de Lyon, CNRS, CRAL, UMR 5574, Saint-Genis-Laval, France}

\authorinfo{Further author information:\\Manon Lizzana E-mail: manon.lizzana@univ-grenoble-alpes.fr\\  Fabien Malbet E-mail: fabien.malbet@univ-grenoble-alpes.fr}

\begin{document} 
\maketitle

\begin{abstract}
Future space astrometry missions such as Theia and an astrometric mode of the Habitable Worlds Observatory (HWO) require detector calibration and instrumental characterization at an unprecedented level to achieve sub-micro-arcsecond precision. This work presents three complementary developments addressing these requirements. First, we report an independent characterization of the GIGAPYX-4600 back-side illuminated CMOS detector, evaluating its linearity, readout noise, dark current, pixel response non-uniformity, defective pixel fraction, and inter-pixel capacitance. The results demonstrate excellent detector performance and confirm its suitability as a candidate for future gigapixel focal planes. Second, we develop an interferometric calibration method based on Young's fringes to measure pixel centroid displacements, enabling the characterization of inter-pixel response variations with an approach compatible with onboard implementation. The method is investigated through numerical simulations and an experimental testbed using the GIGAPYX-4600 detector. Finally, we introduce an astro-calibration framework that jointly estimates stellar astrometric parameters, telescope attitude, plate scale, and optical distortion through a global iterative optimization. Although this calibration approach is still under development, simulations and laboratory activities are underway to validate its performance. Together, these developments contribute to the technological and calibration framework required for the next generation of high-precision astrometric space missions.
\end{abstract}


\section{INTRODUCTION}

Astrometry is the study of the positions, proper motions, and parallaxes of celestial objects. It enables the exploration of many scientific questions about exoplanets, dark matter, stellar clusters, compact objects, and more. Current research, particularly in exoplanet science, demands an accuracy level below 0.3 microarcseconds~\cite{Malbet_2025}. Currently, astrometric detections are limited by the precision on the stellar parameters (positions, proper motions and parallaxes), as achieving such high angular resolution is challenging. This level of precision can only be achieved by future space missions such as Theia or HWO.

Theia is a diffraction-limited telescope project dedicated to astrometry \cite{Malbet_2021}. In May 2025, we submitted Theia to the European Space Agency (ESA) for its M8 call for missions. It will explore many scientific fields, such as exoplanets, dark matter, and compact objects. The telescope has a diameter of $0.8$~m and a focal length of $13$~m. It will operate in visible wavelengths, resulting in a PSF size of approximately $140$~mas at $550$~nm. The FOV is planned to be $0.5$~degrees ($900$~arcmin$^2$), and considering pixels of $4.4$~$\mu$m , it will lead to a $25000 \times 25000$~px  focal plane.

HWO is NASA's next flagship mission \cite{Feinberg_2026}. It will be a serviceable telescope with a wide diversity of scientific objectives, such as searching for habitable worlds, studying galaxy growth, and observing the star formation cycle. The HWO is a good candidate for carrying an astrometric instrument because it provides a stable, space-based telescope with a large aperture of between $6$ and $8$~m, this allows faint sources and small displacements to be observed. The telescope will operate within a wide range of wavelengths between the ultraviolet and infrared spectrum (from $100$ to $2500$~nm). The focal length will be approximately $130$~m, so the PSF size will be about $16$~mas at $550$~nm. The FOV is planned to be $3\times4=12$~arcmin$^2$, considering pixels of $4.4$~$\mu$m, this will lead to a $22000 \times 30000$~px focal plane.

In the context of preparing for future space missions, this work aims to study some of the necessary characterizations and calibrations. Sec.~\ref{sec:carac_det} will present the characterization of the GIGAPYX-4600, a promising focal plane candidate. This section also evaluates the detector's suitability for high-precision astrometry. Sec.~\ref{sec:calib_interfero} presents an interferometric method for measuring pixel centroid displacement. This method allows us to map the inter-pixel response variation of the detector, a crucial aspect for achieving sub-micro-arcsecond precision. Finally, Sec.~\ref{sec:astro_calib} introduces astro-calibration. This consists of determining whether this method can retrieve the optical distortion, plate scale transformation, and star positions.

\section{CHARACTERIZATION OF THE GIGAPYX-4600 CMOS VISIBLE IMAGE SENSOR}
\label{sec:carac_det}

High-precision astrometry demands large-format visible detectors combining excellent photometric performance and low noise. The GIGAPYX-4600, a 46-megapixel back-side illuminated CMOS sensor developed by Pyxalis, represents the first member of a family of detectors intended to build gigapixel focal planes for future space observatories. We independently characterized this detector at IPAG (Grenoble, France)~\cite{Lizzana_2026}.

The GIGAPYX-4600 is a monochrome CMOS detector with an active area of $8320\times5456$ pixels and a pixel pitch of $4.4~\mu$m. Its back-side illuminated architecture provides high quantum efficiency over the visible spectrum, while deep trench isolation is designed to minimize carrier diffusion between adjacent pixels. The detector operates in rolling shutter mode and offers low-gain, high-gain, and dual-gain HDR acquisition modes. Since astronomical observations generally require large full-well capacities, this characterization focused on the low-gain operating mode.

The detector performance was evaluated following the EMVA 1288 standard using the GENEPYX evaluation camera supplied by Pyxalis. The experimental setup consisted of an integrating sphere providing a highly uniform flat-field illumination. Hundreds of bias, dark, and flat-field images were acquired over a range of exposure times to determine the detector gain, linearity, readout noise, dark current, pixel response non-uniformity, defective pixel statistics, and inter-pixel capacitance.

The detector gain was determined from the photon transfer curve, yielding a mean value of $13.9\pm0.8$~e$^-$/ADU. The response remained highly linear over nearly the entire useful dynamic range, with a measured linearity error of only $0.90\pm0.18\%$. The available 12-bit digitization combined with the measured gain resulted in an effective saturation level of approximately $56\ 000$ electrons, providing a sufficiently large dynamic range for astronomical imaging.

Noise characterization demonstrated very good detector performance. The combined readout and quantization noise reached $12\pm3$~e$^-$. This value is slightly lower than the one reported by Pyxalis, the difference is attributed primarily to variations in measurement methodology and readout electronics rather than to intrinsic detector performance.

Dark current measurements required particular attention because of their sensitivity to temperature variations. After allowing the detector to reach thermal equilibrium, the dark current was measured over exposure times up to one minute. The resulting average dark current was $36\pm20$~e$^-$/s at $35^\circ$C. This value is somewhat higher than the manufacturer's published specification of $24\pm20$~e$^-$/s, but discussions with Pyxalis indicated that ongoing improvements of the sensor are expected to significantly reduce dark current in future versions.

Detector uniformity was evaluated through measurements of the Pixel Response Non-Uniformity (PRNU). The overall PRNU measured over the entire detector remained below $1.3\%$, while local measurements yielded values between $1.4\%$ and $1.7\%$ depending on the spatial scale considered. These results indicate a highly uniform detector response, although some large-scale structures originating from detector fabrication and thermal gradients were observed.

A dedicated analysis was also performed to identify defective pixels. Rather than relying on a single threshold, six categories of anomalous behavior were investigated, including hot and cold pixels, noisy response pixels, noisy readout pixels, abnormal dark current, unstable dark current, and abnormal gain. Only $44$ defective pixels were identified across the entire 46-million-pixel array, corresponding to an exceptionally low defect rate of $0.0004\%$. Such a low proportion of defective pixels confirms the excellent manufacturing quality of the detector and minimizes the impact of bad pixels on scientific observations.

One of the most critical aspects for astrometric applications is inter-pixel capacitance (IPC), which introduces electrical coupling between neighboring pixels and may bias stellar centroid measurements~\cite{Rousset_2026}. Using isolated warm pixels observed in dark frames, we characterized the spatial distribution of IPC throughout the detector. The measurements revealed a periodic pattern linked to the detector readout architecture, with each pixel influencing primarily two neighboring pixels. In the worst cases, the coupling remained below $3\%$ of the central pixel signal. Although this level is larger than desired for high-precision astrometry, we demonstrate that IPC can be efficiently corrected using spatially varying deconvolution kernels.

Previous work also discusses the expected behavior of the detector in a space environment~\cite{Michelot_2025}.  Irradiation campaigns performed by Pyxalis demonstrated good tolerance to proton and heavy-ion radiation. While irradiation primarily increases dark current and temporal noise, other key characteristics—including linearity, saturation level, and pixel response uniformity—remain essentially unchanged. Furthermore, no latch-up events were observed during heavy-ion testing, supporting the detector's suitability for long-duration space missions.

\begin{table}[ht]
\caption{GIGAPYX-4600 characterization, comparison between IPAG and Pyxalis~\cite{Lizzana_2026}} 
\label{tab:comparsion_IPAG_Pyxalis}
\begin{center}       
\begin{tabular}{|l|l|l|} 
\hline
\rule[-1ex]{0pt}{3.5ex}  Property & Pyxalis & IPAG\\
\hline\hline
\rule[-1ex]{0pt}{3.5ex}  read out noise & 16 e$^-$ & 12 e$^-$\\
\hline
\rule[-1ex]{0pt}{3.5ex}  linearity error & 2\% & 0.90\%\\
\hline
\rule[-1ex]{0pt}{3.5ex}  response non uniformity & $\leq$ 2\% & 1.3\%\\
\hline
\rule[-1ex]{0pt}{3.5ex}  dark current (at $35^\circ$C) & 24 e$^-$/s & 36 e$^-$/s\\
\hline 
\rule[-1ex]{0pt}{3.5ex}  saturation charge & $>$50\,ke$^-$ & 56\,ke$^-$\\
\hline 
\rule[-1ex]{0pt}{3.5ex}  percentage of defective pixels &$\leq$ 0.1\%& $0.0004\%$ \\
\hline 
\rule[-1ex]{0pt}{3.5ex}  inter-pixel capacitance &- & $\leq$ 3\% \\ 
\hline 
\end{tabular}
\end{center}
\end{table}

Table~\ref{tab:comparsion_IPAG_Pyxalis} shows a comparison of the results of our investigation with the Pyxalis results given in their datasheet. Comparison with the manufacturer's specifications shows excellent agreement for nearly all measured parameters. The detector exhibits lower-than-expected readout noise, improved linearity, comparable pixel response uniformity, and an extremely low defective pixel rate. The only noticeable discrepancy concerns the dark current, which remains slightly higher than the manufacturer's measurements but is expected to decrease in future detector revisions.

Overall, the independent characterization confirms that the GIGAPYX-4600 fulfills the main performance requirements expected for future high-precision astrometric instruments. The detector combines excellent linearity, low readout noise, high response uniformity, very few defective pixels, and manageable inter-pixel capacitance. The present results establish the GIGAPYX-4600 as a strong technological candidate for future modular gigapixel focal planes envisioned for missions such as HWO and Theia.

\section{INTERFEROMETRIC CALIBRATION OF THE PIXEL POSITIONS}
\label{sec:calib_interfero}

A detector is made up of individual pixels, each of which has its own structure and spatial response to light. The centroid of a pixel can be defined as the barycenter of the spatial response. Since the intra-pixel response is not a perfect rectangular function and varies from pixel to pixel, the pixels' centroids do not align perfectly in rows and columns. The displacements between the real centroids and the ideal centroids (perfect positions: if the pixels were well aligned and if the intra-pixel responses were rectangular functions) are called $(\delta_x,\delta_y)$. Measuring these displacements allows for mapping the inter-pixel response variation of the detector.

Different methods, like the scan with an electron beam or a light source, allow us to estimate the intra-pixel response. However, scanning a large number of pixels is impractical because it is very slow and expensive. This is why our work focuses on measuring pixel centroids using the interferometric method because it is the only method that can easily be carried out onboard.

A method has been imagined to measure pixel centroids. Its goal is to be simple and fast to carry out on a space mission. It relies on a laser source that is split into two fibers to illuminate the focal plane. This creates Young fringes on the pixels. These fringes can be moved across the detector by shifting the phase of light in one fiber relative to the other. Each pixel detects sinusoidal variations in light intensity. This variation reveals the pixel's relative position in the direction perpendicular to the fringes. Another pair of fibers enables measurement in a second direction, providing a two-dimensional estimation of pixel locations.

Previous studies have shown the performance of this method on $80\times80$ pixel CCD detectors with a $24~\mu$m pitch. A testbed was developed in the United States that first achieved an average precision of $\es{1.7}{-5}$~px on a $10\times10$ pixel area \cite{Nemati_2011} and then an accuracy of $\es{1}{-4}$~px on individual pixels \cite{Shao_2023}. A similar testbed was developed in France and achieved a precision of $\es{4}{-4}$~px on individual pixels and $\es{6}{-5}$~px on a PSF photocenter \cite{Crouzier_2016}.

We developed an algorithm based on those used in previous studies. The system can be modeled using free parameters, including $(\delta_x,\delta_y)$ which we will attempt to estimate. These parameters must be fitted using least squares minimization between the models and the data. Since this problem is nonlinear and contains different types of parameters (spatial and temporal), an iterative process seems optimal. First, a spatial fit is performed on each frame to constrain the temporal parameters (the spatial parameters are kept constant), and then a temporal fit is performed on each pixel to constrain the spatial parameters (the temporal parameters are kept constant).

Synthetic measurements were computed to simulate the focal plane and test this algorithm. It allows different configurations to be explored by varying parameters such as the length of the baseline, the size of the focal plane, the number of frames, or the evolution of the temporal phase. More details about the algorithm implementation and the results obtained through simulation are provided in Rousset et al. (2026)~\cite{Rousset_2026}.

Moreover, to demonstrate the ability of this method to calibrate pixel centroids in large matrices with small pixels and to show the performance of the described algorithm, a testbed was set up at IPAG (Grenoble, France). An HeNe laser source ($632$~nm) is divided into two fibers by a splitter. A LiNbO3 electro-optic modulator from Jenoptik modulates the phase between the two beams. A fiber switch then allows one to choose the direction and baseline between the outputs. The diffraction pattern is recorded by the GIGAPYX-4600 detector at a distance of $40$ centimeters. More details about the testbed and the results obtained through real data are provided in Rousset et al. (2026)~\cite{Rousset_2026}.

\section{ASTRO-CALIBRATION}
\label{sec:astro_calib}

The measured position of a star is a linear distance on the focal plane; however, the astronomically interesting position is an angular distance in a celestial coordinate system. Many transformations occur between these two positions, such as deformation due to optical transformation, homothety between pixels and arcseconds, translation and rotation due to the telescope's line of sight, and projection between linear and spherical coordinates. All these effects must be understood and corrected or calibrated in order to detect an astrometric signal.

Five main types of methods have been developed in the literature to calibrate the effects described below: the reference catalog match~\cite{Libralato_2024}, the dithered observations~\cite{Anderson_2003}, the diffractive pupil~\cite{Guyon_2012}, the artificial sources~\cite{Rodeghiero_2019} and the joint calculation~\cite{Lindegren_2012}. In the context of high-precision astrometry, the adopted solution must be able to handle sub-micro-arcsecond accuracy and temporal evolution. The joint calculation method seems to be the most promising in the context of high-precision astrometry because it is not very demanding: the stars in the FOV are directly used as metrology sources, and Gaia shows that it can carry out precise measurements \cite{gaia_2023}.

\begin{figure}[t]
   \begin{center}
   \includegraphics[width=1\linewidth]{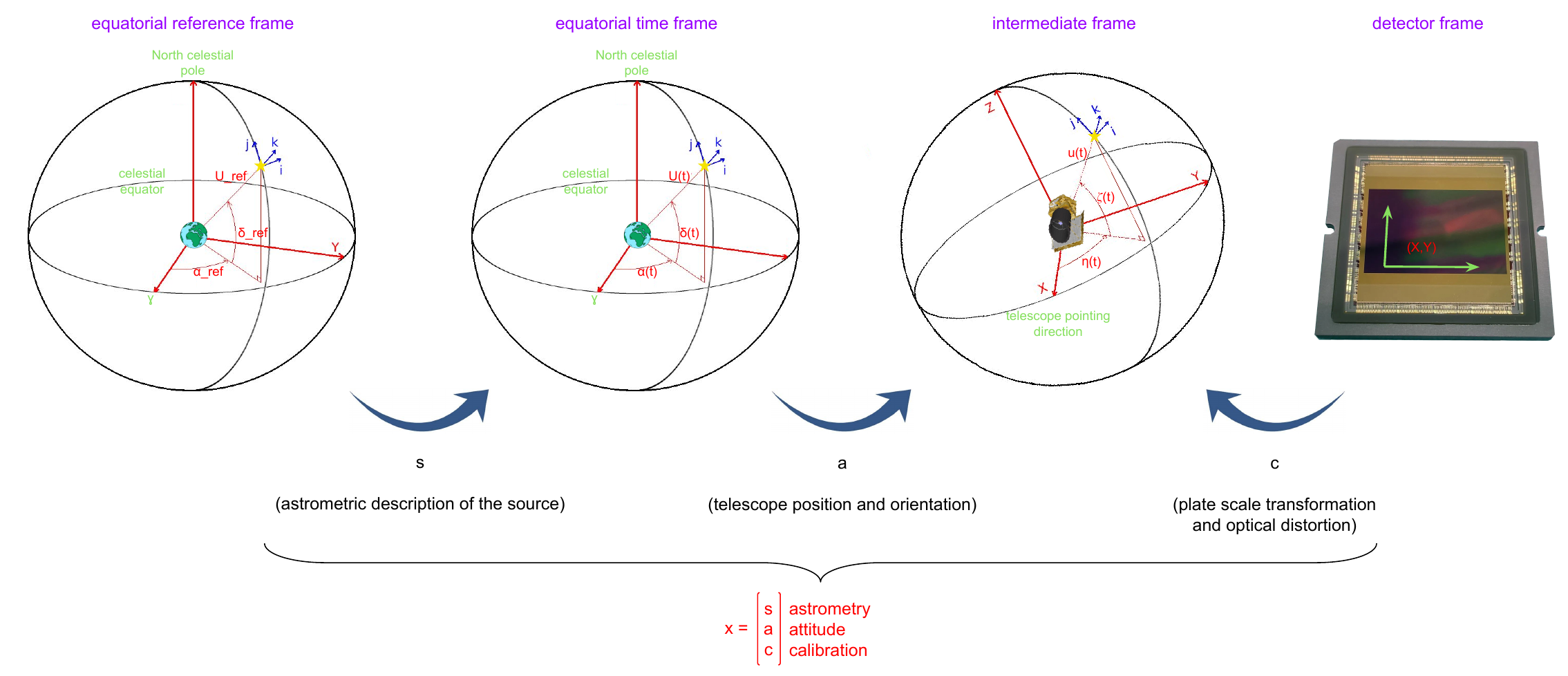}
   \end{center}
   \caption{Illustration of the astro-calibration method. The four reference frames are represented. The three models (\textit{s}, \textit{a}, and \textit{c}) that link them and constitute the parameter vector $x$ are also represented.}
   \label{fig:schema_astro_calib}
\end{figure}

We developed a method inspired from Lindegren et al. (2012)~\cite{Lindegren_2012} to calibrate the astrometry, the telescope attitude, the plate scale transformation and the optical distortion, all at ones. First, it is necessary to define four different frames as insulated in Fig.~\ref{fig:schema_astro_calib}. The first one is called the \textit{equatorial reference frame}, it describes the position of the stars in terms of right ascension and declination at a specific time. The second one is calle the \textit{equatorial time frame}, it also describes the position of the stars in terms of right ascension and declination, however, these values evolve over time due to the proper motions of the stars. The third one is called the \textit{intermediate frame}, it is a reference frame linked with the telescope (the x-direction is aligned with the pointing direction). Stellar positions are described with angles $\left(\eta(t),\zeta(t)\right)$. The fourth one is called the \textit{detector frame}, it describes stellar positions as distances in pixels on the focal plane.

These reference frames are connected by three models: \textit{s}, \textit{a}, and \textit{c}, each of which relies on a set of corresponding parameters. The \textit{s} model describes astrometry and links the \textit{equatorial reference frame} and the \textit{equatorial time frame}. It consists of a linear model that describes the stars' trajectories over time. The associated parameters are the positions and proper motions of each star. The \textit{a} model describes the telescope's attitude, (pointing directions and rotations over time), and links the \textit{equatorial time frame} and the \textit{intermediate frame}. It consists of 3D rotations modeled by quaternions that describe the telescope's orientation for each observation. The associated parameters are triplets of angles that determine the telescope's attitude. The \textit{c} model describes calibration: the plate scale transformation and optical distortion, and linking the \textit{detector frame} and \textit{intermediate frame}. It consists of two coupled 2D polynomials. The associated parameters are the coefficients of the polynomials for each observation. All of these parameters can be grouped into a vector called $x=(s,a,c)$.

Approximations of the previously described parameters are possible. For example, an approximation of the astrometric parameters can be obtained from the Gaia catalog. The telescope attitude can be approximated using gyroscopes and guiding stars. The calibration can be estimated from ground measurements before flight. However, these approximations would not allow us to reach sub-micro-arcsecond precision. For this reason, all of these parameters are considered unknown and must be fitted.

For a given observation $i$ and a given star $j$, 2 couples of coordinates $(\eta_{ij},\zeta_{ij})$ in the intermediate frame can be calculated (one using \textit{s} and \textit{a} and another one using only \textit{c}):

\begin{equation}
    \begin{cases}
        \eta_{\mathrm{obs}\,ij}(c)=\left(X_{ij}+\sum_m \sum_n c_{\eta imn} X_{ij}^m Y_{ij}^n\right)/F \\
        \zeta_{\mathrm{obs}\,ij}(c)=\left(Y_{ij}+\sum_m \sum_n c_{\zeta imn} X_{ij}^m Y_{ij}^n\right)/F
    \end{cases}
\end{equation}

\begin{equation}
    \begin{cases}
        \eta_{\mathrm{mod}\,ij}(s,a)=atan2(u_{y\,ij},u_{x\,ij}) \\
        \zeta_{\mathrm{mod}\,ij}(s,a)=atan2(u_{z\,ij},\sqrt{u_{x\,ij}^2+u_{y\,ij}^2})
    \end{cases}
\end{equation}

where the $(X_{ij},Y_{ij})$ are the photocenter of the stars measured in the detector frame, $F$ is the focal length of the telescope, and $\vec{u_{ij}}=\left(u_{x\,ij},u_{y\,ij},u_{z\,ij}\right)$ is a unit vector describing the direction telescope-star.

We can defined a cost function $Q$ which is a weighted distance between theses two couples of coordinates:
\begin{equation}
    Q(x) = \sum_{ij} \left[(\eta_{\mathrm{obs}\,ij}-\eta_{\mathrm{mod}\,ij})^2+(\zeta_{\mathrm{obs}\,ij}-\zeta_{\mathrm{mod}\,ij})^2\right] W_{ij}
    = \sum_{ij} \left[R_{\eta\,ij}(x)^2+R_{\zeta\,ij}(x)^2\right] W_{ij}
    = \sum_{ij} R_{ij}(x)^2 W_{ij}
\end{equation}

where $W_{ij}$ is a weight that may depend on the observation and star considered.

\begin{figure}[t]
   \begin{center}
   \includegraphics[width=0.34\linewidth]{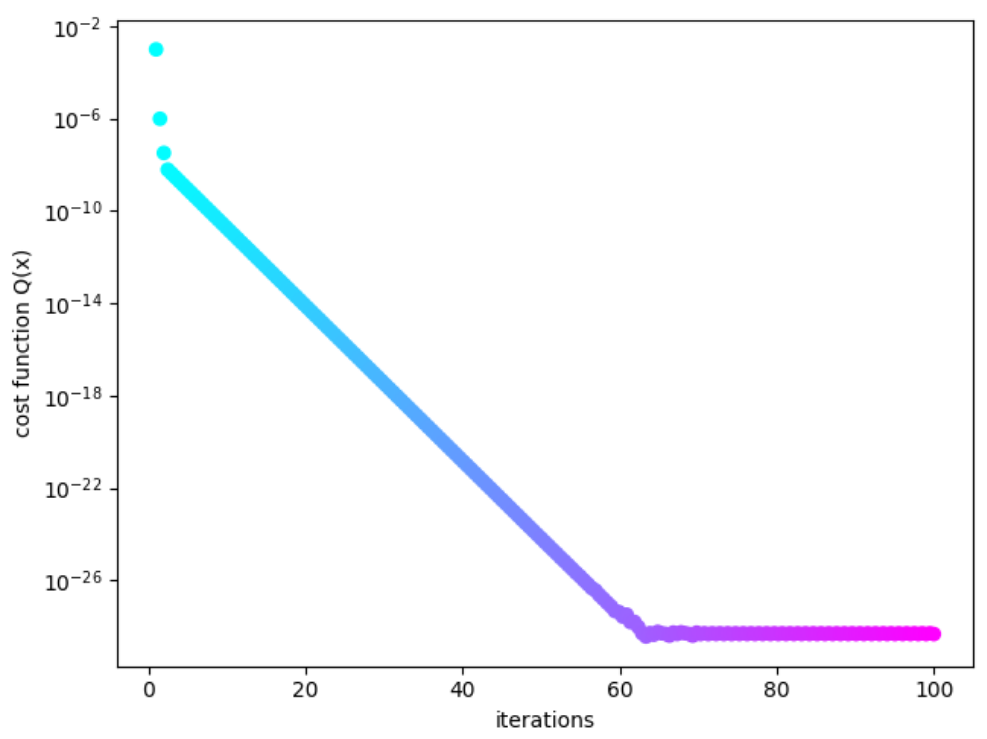}
   \includegraphics[width=0.65\linewidth]{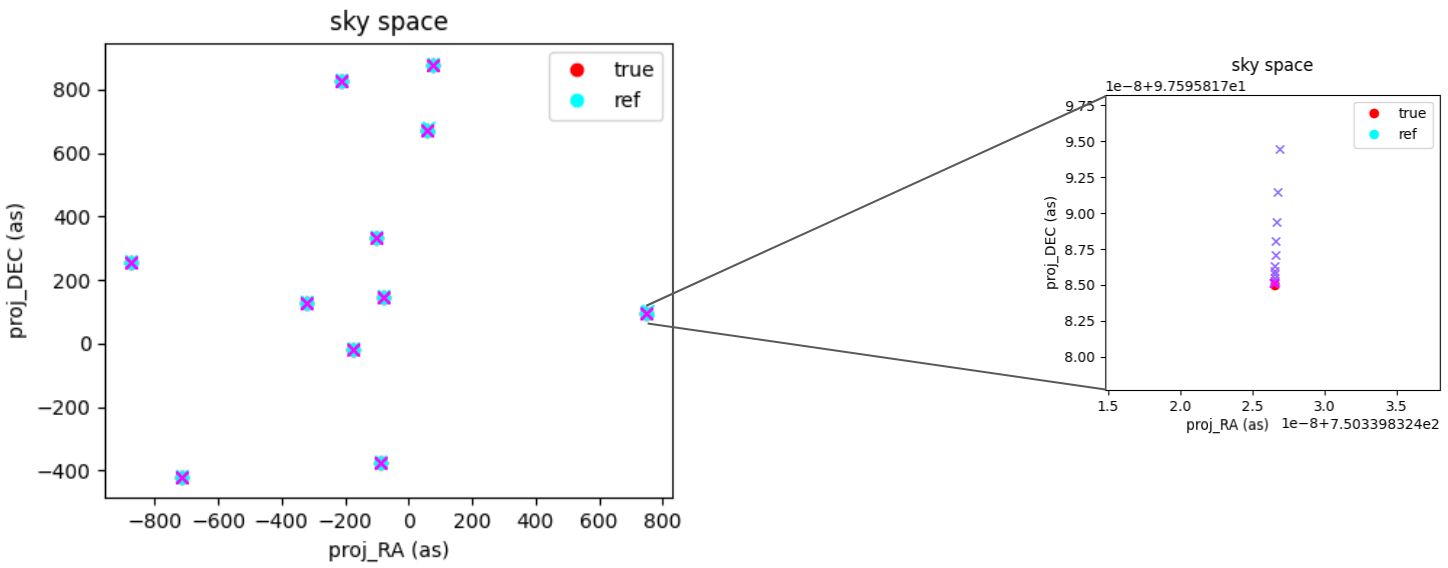}
   \end{center}
   \caption{Left: example of the decrease in the cost function $Q(x)$ across the iterations. Right: example of a simulated field of view composed of $11$ stars. The evolution of the parameters \textit{a} representing the stars' positions is also shown by the movement of the cross from the 'ref' point (the initial guess, starting point) to the 'true' point (the true position, which was injected at the beginning of the simulation).}
   \label{fig:cost_func_params_a}
\end{figure}

We are searching for the parameter vector x that minimizes the cost function: $x=argmin_x\left(Q(x)\right)$. In practice, the $x$ vector is divided in three parts: $s$, $a$ et $c$, and the minimization is performed iteratively by blocks. The minimizations of $s$ and $a$ are carried out with a Levenberg-Marquardt algorithm because they are nonlinear problems. By contrast, $c$ is a linear problem and constitutes a classical weighted least squares problem. Fig.~\ref{fig:cost_func_params_a} shows an example of the behavior of the cost function and the convergence of the parameters representing the star positions.

For now, we are creating a simulation to obtain synthetic measurements and model the telescope's attitude and optical distortion. We have also carried out the first tests of this method, but it is an ongoing project and no robust results are currently available.

We also setup a testbed at IPAG (Grenoble, France) to test the performance of the astro-calibration. The main objective is to demonstrate in a laboratories that optical distortion, plate scale transformation, and star positions can be retrieved. The goal is not to create a realistic replica with a telescope model or an entire focal plane, which would be very complex, rather, the goal is to provide a basic proof of concept of the method and explore its limits under simple conditions.

\section{CONCLUSION}

High-precision astrometry at the sub-micro-arcsecond level requires both high-performance detectors and advanced calibration strategies. In this work, we presented three complementary developments addressing these challenges. The independent characterization of the GIGAPYX-4600 CMOS detector confirms that its performance satisfies the main requirements for future astrometric missions, with excellent linearity, low readout noise, high response uniformity, and a very low defective-pixel rate. We also presented an interferometric method for measuring pixel centroid displacements, together with dedicated simulation and laboratory developments aimed at demonstrating its applicability to large-format CMOS detectors. Finally, we introduced a global astro-calibration approach that jointly estimates astrometric, attitude, and instrumental calibration parameters. While this work is still in progress, ongoing simulations and experimental validation will assess its capability to recover optical distortion, plate scale, and stellar positions with the accuracy required for future missions such as Theia and HWO.

\acknowledgments
 
The work has been also partially supported by the LabEx FOCUS by the French National Research Agency through the grant ANR-11-LABX- 0013 and within the framework of the "France 2030” program by the grant ANR-15-IDEX-02. The authors acknowledge financial support from the Centre national d’études spatiales (CNES), France (ROR: \url{https://ror.org/04h1h0y33}), thanks to the Research and Technology (R\&T) and mission preparation programs. Manon Lizzana acknowledges the support of her PhD grant from CNES and Pyxalis.
This research has made use of the Astrophysics Data System, funded by NASA under Cooperative Agreement 80NSSC21M0056.

\bibliography{report} 
\bibliographystyle{spiebib} 

\end{document}